# Geo-Based Scheduling for C-V2X Networks

Rafael Molina-Masegosa, Miguel Sepulcre and Javier Gozalvez

*Abstract*—Cellular Vehicle-to-Everything (C-V2X) networks can operate without cellular infrastructure support. Vehicles can autonomously select their radio resources using the sensing-based Semi-Persistent Scheduling (SPS) algorithm specified by the Third Generation Partnership Project (3GPP). The sensing nature of the SPS scheme makes C-V2X communications prone to the well-known hidden-terminal problem. To address this problem, this paper proposes a novel geo-based scheduling scheme that allows vehicles to autonomously select their radio resources based on the location and ordering of neighboring vehicles on the road. The proposed scheme results in an implicit resource selection coordination between vehicles (even with those outside the sensing range) that reduces packet collisions. This paper evaluates analytically and through simulations the proposed scheduling scheme. The obtained results demonstrate that it reduces packet collisions and significantly increases the C-V2X performance compared to when using the sensing-based SPS scheme.

## I. INTRODUCTION

The 3GPP published in Release 14 a new standard to support V2X communications [1]. This standard is known as C-V2X (or Cellular-V2X, Long Term Evolution V2X, LTE-V2X, LTE-V), and supports direct V2V (Vehicle-to-Vehicle) communications with the PC5 interface (also referred to as sidelink). 3GPP Release 15 evolves C-V2X, and defines enhanced use cases [2]. 3GPP has currently an open study item under Release 16 to define the evolution of C-V2X [3].

C-V2X is the first cellular standard that allows direct V2V communications. C-V2X communications are managed by the cellular infrastructure when operating under mode 3, and several algorithms have been proposed in the literature for radio resource allocation (e.g. [4] and [5]). C-V2X mode 3 can benefit from the centralized control of the radio resource allocation, but it is only valid under cellular coverage. On the other hand, C-V2X mode 4 can operate with and without cellular coverage. C-V2X mode 4 allows vehicles to autonomously select their radio resources. C-V2X mode 4 is highly relevant since V2V safety applications cannot depend on the availability of cellular coverage. C-V2X mode 4 defines a sensing-based Semi-Persistent Scheduling (SPS) algorithm that all vehicles implement to select their radio resources. The SPS scheme is sensing-based, so vehicles use the algorithm to estimate which resources are available. The sensing nature of the SPS algorithm results in that C-V2X mode 4 can be challenged by the hidden terminal problem: if two vehicles cannot sense each other, they can select and simultaneously transmit using the same radio resource which results in packet collisions [6]. Additional C-V2X mode 4 challenges include half-duplex operation and scalability. Standardized C-V2X equipment can only operate under Half Duplex mode [7], so vehicles cannot transmit and receive at the same time. Consequently, vehicles cannot overhear all other vehicles transmitting at the same time even if they transmit using different sub-channels. The negative effects resulting from hidden terminal and half-duplex operation, together with the bandwidth limitations (particularly relevant when considering bandwidth-demanding automated applications), influence the capacity of C-V2X networks to effectively support large number of vehicles. Mechanisms are hence necessary to control the load and ensure the scalability of the C-V2X network while maintaining high reliability levels.

Different studies have proposed options to improve the performance of C-V2X mode 4 and its SPS scheme. For example, [8]-[11] analyze and optimize the parameters of the SPS algorithm that are not fixed by the 3GPP standard. [12]-[15] propose modifications or extensions to the SPS algorithm but maintain most of its functionality. For example, [12] proposes that vehicles execute earlier the radio resource reselection process so that they can inform other vehicles in advance of the resources they will utilize for their following transmissions. A related objective is sought in [13] where authors propose that vehicles should inform other vehicles about the number of packets that are going to be transmitted using the same radio resources. [14] proposes that vehicles only reserve resources for the more frequent (and smaller) packets. Less frequent (and larger) packets should be transmitted without reserving resources (one shot transmissions). [15] modifies SPS so that higher weight is given to the most recent signal level measurements when selecting new resources. [16] proposes a power control algorithm to reduce interference when

Copyright (c) 2019 IEEE. Personal use of this material is permitted. However, permission to use this material for any other purposes must be obtained from the IEEE by sending a request to pubs-permissions@ieee.org.

This work was supported in part by the *Conselleria de Educación, Investigación, Cultura y Deporte* of *Generalitat Valenciana* through the project AICO/2018/A/095, the Spanish Ministry of Economy and Competitiveness and FEDER funds under the project TEC2017-88612-R, and grant PEJ-2014-A33622. Rafael Molina-Masegosa, Miguel Sepulcre, and Javier Gozalvez are with the UWICORE Laboratory, Universidad Miguel Hernandez de Elche (UMH), Spain. E-mail: rafael.molinam@umh.es, msepulcre@umh.es, j.gozalvez@umh.es.

the channel load is high. A different alternative to control the channel load is packet dropping, which is evaluated in [17]. All these proposals improve the performance of SPS and C-V2X mode 4. However, they are still affected by the challenges previously discussed, and in particular by the hidden-terminal problem that limits their gains. Combatting the hidden terminal problem requires vehicles to be aware of which radio resources are used by vehicles beyond their sensing range. This is the objective of this study that proposes a novel geo-based scheduling scheme for C-V2X Mode 4 that combats the hidden terminal challenge.

The proposed geo-based scheduling scheme selects radio resources based on the location of vehicles and their ordering on the road. The ordering is estimated using the location and speed information received in the vehicles' beacons. Beacons are referred to as CAMs (Cooperative Awareness Messages) in European Telecommunications Standards Institute (ETSI) standards and BSMs (Basic Safety Messages) in Society of Automotive Engineers (SAE) standards. All vehicles autonomously select their radio resources following the geographical ordering of the vehicles on the road. This coordinated process reduces packet collisions even with vehicles beyond the sensing range. The performance of the proposed scheme is analyzed analytically and through network simulations in highway scenarios. The results demonstrate that the geo-based scheduling scheme reduces packet collisions and significantly increases the C-V2X Mode 4 performance compared to when using the sensing-based SPS scheme.

## II. C-V2X COMMUNICATIONS

C-V2X mode 4 vehicles autonomously select their radio resources or sub-channels to transmit their data and control information [18]. A sub-channel is defined in C-V2X as a group of Resource Blocks (RBs) in the same 1ms sub-frame (Fig. 1). C-V2X divides its 10MHz (or 20MHz) channels into 1ms sub-frames and RBs of 180 kHz each. The data is transmitted over Transport Blocks (TBs), and the control information is transmitted in Sidelink Control Information (SCI) messages. Each TB contains a full packet (e.g. a beacon), and has an associated SCI that must be transmitted in the same sub-frame and occupies 2 RBs. The SCI includes important information that must be correctly received to decode the associated TB, and that includes among others: the Modulation and Coding Scheme (MCS) used to transmit the TB, the sub-channels occupied by the TB, and the resource reservation interval. This interval indicates how often the vehicle intends to use the selected sub-channels. It is equal to 100ms if vehicles transmit 10 beacons or packets per second (i.e. 10pps); 50ms for 20pps; and 20ms for 50pps. Vehicles reserve the selected sub-channels for *Reselection Counter* consecutive transmissions so that other vehicles can accurately estimate which sub-channels are occupied. *Reselection Counter* is a random number selected between: 5 and 15 for 10pps; 10 and 30 for 20pps; and 25 and 75 for 50pps. It is decremented by one after each packet transmission, and the vehicle must select with probability (1-$P$) new sub-channels when *Reselection Counter* is equal to zero. Following [8], $P$ is configured in this study equal to 0.

Vehicles select their sub-channels using the sensing-based SPS scheme defined in 3GPP Release 14 [18]. Vehicles are required to continuously sense the transmissions in all sub-channels to avoid selecting sub-channels that are used by other vehicles. To select its sub-channels, a vehicle must first identify the Selection Window (SW) where it can search for the candidate sub-channels. The SW is illustrated in Figure 1, and is defined as the time period between the time $T$ at which the the vehicle wants to reserve a sub-channel and $T$ plus the maximum transmission latency. This latency is defined in the 3GPP standard [18] and is equal to 100ms, 50ms and 20ms when vehicles transmit 10pps, 20pps and 50pps, respectively. Within SW, the vehicle identifies all the Candidate Single-Subframe Resources (CSRs). A CSR is a group of adjacent sub-channels within the same sub-frame where the SCI+TB to be transmitted fits. The vehicle then creates a list $L_1$ with all the CSRs in SW except those for which the vehicle: (1) has received in the previous 1000 sub-frames an SCI from another vehicle indicating that it will utilize this CSR at the same time that the vehicle will need it to transmit any of its next packets; and (2) has measured an average Reference Signal Received Power (RSRP) in the associated TB higher than a pre-defined threshold (equal to -120dBm in this study). If $L_1$ does not include at least 20% of all CSRs in SW, the vehicle increases the RSRP threshold by 3dB and generates a new list $L_1$. This process is repeated until $L_1$ includes at least 20% of all CSRs in SW. The vehicle then creates a second list $L_2$ that includes the CSRs from $L_1$ with the lowest average RSSI (Received Signal Strength Indicator). The average RSSI is computed over all its RBs and over the previous $T_{CSR}$-$j$·$T_{IPI}$ sub-frames ($j \in \mathbb{N}$), where $T_{IPI}$ is equal to 100, 50 and 20 when vehicles transmit 10pps, 20pps and 50pps respectively (Fig. 1). $j$ is defined as $1 \leq j \leq 10$ for 10pps, $1 \leq j \leq 20$ for 20pps, and $1 \leq j \leq 50$ for 50pps. The total number of CSRs in $L_2$ must be equal to 20% of all CSRs in SW. Finally, the vehicle randomly chooses one of the CSRs in $L_2$, and utilizes it for the next *Reselection Counter* packet transmissions. New resources must be selected if a packet to be transmitted does not fit in the CSR previously reserved. It should be noted that current 3GPP standards only consider half duplex operation for C-V2X radios, i.e. a vehicle cannot transmit and receive at the same time [7].

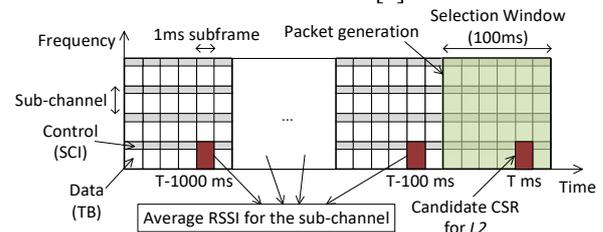

Fig. 1. C-V2X mode 4 channelization and scheduling for 10 pps [19].

## III. GEO-BASED SCHEDULING

The objective of the geo-based scheduling is to reduce packet collisions by coordinating the selection of sub-channels between vehicles, even between vehicles that cannot sense each other. To this aim, vehicles exploit the geographical location exchanged through beacons to form a virtual queue. The order of the vehicles in the virtual queue is used by the proposed geo-based scheduling to deterministically select the sub-channels that will be used by each vehicle to transmit. This deterministic selection has been designed to maximize the distance between two vehicles using the same sub-channel, and therefore minimize packet collisions.

To achieve its objectives, the proposed scheme organizes the use of sub-channels into *Pools* that are common to all vehicles. A *Pool* is defined as the set of sub-channels included in all the sub-frames within a given time period. This time period is here set up equal to the packet transmission interval of beacons. A *Pool* has a total of $N$ sub-channels:

$$N = SF \cdot SC = \frac{1000}{\lambda} \cdot SC \quad (1)$$

where $SF$ is the number of 1ms sub-frames in the *Pool* and $SC$ is the number of sub-channels per sub-frame. $SF$ is set equal to $1000/\lambda$, where $\lambda$ is the packet transmission frequency in pps (i.e. the inverse of the packet transmission interval). Fig. 2 illustrates the organization of a *Pool*. All vehicles are time-synchronized at the *Pool* level since they are synchronized at the sub-frame level (following 3GPP specifications [20]) and sub-frames are numbered using the Direct Frame Number (DFN) [21]; the DFN can be derived even if there is no cellular coverage. This does not imply that vehicles generate their beacons at the same time. The proposed scheduling scheme is completely distributed and is executed by all vehicles at the end of the time period that defines a *Pool*. It operates as follows:

1. All vehicles estimate at the end of a *Pool* the location of their neighboring vehicles. To this aim, vehicles use the location, speed and *TimeStamp* data transmitted by neighboring vehicles in their last beacons.
2. Each vehicle builds a virtual queue and orders all its neighboring vehicles in the queue based on their geographical location. The position of a vehicle in the queue is represented by an integer number referred to as *PosIndex*. *PosIndex* can take values between 0 and $N$-1 where $N$ is equal to the number of sub-channels in a *Pool*. Vehicles attach their *PosIndex* to their beacons[1].
3. To compute their *PosIndex*, vehicles take into account the ordering of vehicles in the queue and the *PosIndex* information transmitted by neighboring vehicles. If a vehicle overhears that its preceding vehicle has a *PosIndex* equal to $X$, it chooses a *PosIndex* equal to $X+1$. If a vehicle does not detect any neighboring vehicle, it maintains its last *PosIndex* value.
4. Vehicles select their sub-channel(s) for transmission based on their *PosIndex*. Packet collisions can then be avoided if all vehicles correctly estimate their *PosIndex*. Such collisions are unavoidable if there are more vehicles than sub-channels in a *Pool*. However, if this is the case, the geo-based scheduling scheme maximizes the distance between vehicles that share the same sub-channels.

TABLE I. VARIABLES

| Variable | Definition |
|---|---|
| $\beta$ | Traffic density (vehicles per meter) |
| $\delta_{HD}$, $\delta_{SEN}$, $\delta_{PRO}$, $\delta_{COL}$ | Probability of not correctly receiving a packet due to HD, SEN, PRO and COL errors, respectively |
| $\Delta PI_{t,r}$ | Difference between PosIndex of vehicles $v_t$ and $v_r$ |
| $d_{t,u}$ | Distance between $v_t$ and its preceding vehicle $u$ |
| $\lambda$ | Packet transmission frequency (pps) |
| $2M$ | Length of the Random Transmission Window |
| $N$ | Number of sub-channels in a Pool |
| $N_P$ | Number of sub-channels occupied by the transmitted packet |
| $PI_u$ | PosIndex of vehicle $u$ |
| $PR$ | Center of the Random Transmission Window |
| $P_{SEN}$ | Sensing power threshold |
| $SC$ | Number of sub-channels in a sub-frame |
| $sc(PI)$ | First sub-frame in the sub-frame $sf(PI)$ |
| $SF$ | Number of 1ms sub-frames in a Pool |
| $sf(PI)$ | Sub-frame in the Pool corresponding to PosIndex $PI$ |
| $U$ | Number of detected preceding vehicles |
| $\mu$, $\eta$ | Weighting parameters for PosIndex estimation |
| $w$ | Number of transmissions in normal mode between two consecutive transmissions in random mode |
| $w_{min}$, $w_{max}$ | Minimum and maximum value of $w$ |

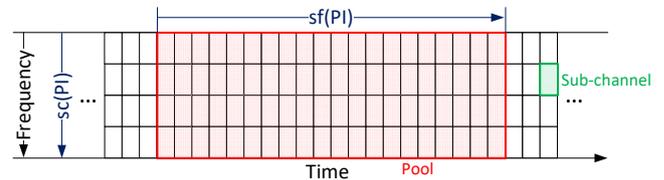

Fig. 2. *Pool* organization for $\lambda$=50pps ($SF$=20), $SC$=4 and $N$=80.

### A. Location estimation and virtual queues

All vehicles estimate the location of their neighboring vehicles at the end of each *Pool* and using the information included in their last beacon. The location of a neighboring vehicle is estimated as the sum of the location included in its last beacon[2] plus the distance traveled since such last beacon was generated[3]. Similarly, each vehicle estimates its own location at the end of a *Pool* considering the location it included in its last beacon and the distance travelled since the last beacon was generated. Vehicles then order their neighboring vehicles in a virtual queue based on their estimated geographical location. The process followed to estimate the location of vehicles is robust against low GNSS location accuracies, and guarantees that all vehicles estimate the same location for a given neighboring vehicle. Consequently, all vehicles equally order their neighboring vehicles in their virtual queues, and will not select the same *PosIndex*. In addition, the scheduling scheme is robust against errors in the ordering of vehicles in the queue since all vehicles identify the same order.

### B. PosIndex estimation

Vehicles use the virtual queue to estimate their *PosIndex* considering only their preceding vehicles. If a vehicle does not detect any preceding vehicle, it maintains its last *PosIndex* value. We denote as $U$ the number of detected preceding vehicles. Vehicles are numbered from 1 to $U$ with vehicle $v_1$ being the closest preceding vehicle to the vehicle $v_t$ that is computing its *PosIndex*, and vehicle $v_U$ the farthest. If the *PosIndex* of vehicle $u$ is $PI_u$ then vehicle $u$-1 should have a *PosIndex* equal to $PI_u$+1. The maximum value for the *PosIndex* is $N$-1, so the vehicle that follows the vehicle with *PosIndex* equal to $N$-1 selects a *PosIndex* equal to 0. If all vehicles were perfectly ordered, vehicle $v_t$ should just check the *PosIndex* of preceding vehicle $v_1$ and select its *PosIndex* equal to $PI_1$+1. However, there can be errors in the estimation of the *PosIndex*, and hence $v_t$ has to check the *PosIndex* of all its preceding vehicles. $v_t$ computes then its *PosIndex* as:

$$PI_t = \arg \max_{PI} \sum_{u=1}^{U} \Phi(PI, PI_u + u) \quad (2)$$

where

$$\Phi(X,Y) = \begin{cases} 1 & if \quad X = Y \\ 0 & if \quad X \neq Y \end{cases} \quad (3)$$

---

[1] This requires an overhead of 8 to 12 bits per beacon depending on the number of sub-channels per sub-frame. This overhead has been taken into account in our evaluation.

[2] For robustness, we only consider vehicles from which at least one of their last ten beacons has been received.

[3] This distance is equal to the multiplication of the time elapsed since the last beacon was generated and the vehicles' speed (included in the beacons). The beacons also include the parameter *TimeStamp* that indicates the time at which the last beacon was generated.

Equation (2) weights equally the *PosIndex* of all the preceding vehicles for robustness. However, it is not optimal to rapidly update the *PosIndex* values when there is a change in the ordering of vehicles in the queue. This is illustrated in Fig. 3 where $v_t$ needs to compute its *PostIndex*. Let's suppose that the vehicle has detected three preceding vehicles in its sensing range, and their *PosIndex* are equal to 9, 15 and 16, respectively. The vehicle computing the *PosIndex* should select a *PosIndex* equal to 12 since the correct ordering of the preceding vehicles should be 9, 10 and 11 (vehicle 9 has a preceding vehicle with *PosIndex* equal to 8). However, vehicles with *PosIndex* equal to 15 and 16 have not yet modified their *PosIndex*, and (2) results in that $v_t$ selects a *PosIndex* equal to 17. This vehicle will only select the correct *PosIndex* (i.e. 12) when vehicle 15 updates its *PosIndex*. The *PosIndex* could be updated faster if it was chosen only as a function of the *PosIndex* of the farthest vehicle. However, this will make the scheme very unreliable if a single vehicle incorrectly estimates its *PosIndex*. There is then a trade-off between robustness and speed to update the *PosIndex* values, and we propose to address it by replacing (2) with (4):

$$PI_t = \arg \max_{PI} \sum_{u=1}^{U} \Phi(PI, PI_u + u) \cdot (\mu + \eta \cdot d_{t,u}) \quad (4)$$

where $d_{t,u}$ is the distance between $v_t$ and each of its preceding vehicles. $\mu$ and $\eta$ are weighting parameters. A high $\mu/\eta$ increases the robustness of the *PosIndex* estimation, while a low $\mu/\eta$ increases the speed to update the *PosIndex* when there are changes in the ordering. $\mu$ and $\eta$ are set equal to 10 and 0.1 to achieve a compromise between robustness and speed.

### C. Sub-channel selection

Vehicles select their sub-channel(s) from the *Pool* based on their *PosIndex*. The organization of a Pool is illustrated in Fig. 2. We number the sub-frames in a *Pool* from 0 to *SF*-1 and the sub-channels in a frame from 0 to *SC*-1. A vehicle with *PosIndex* equal to *PI* selects a sub-channel in the sub-frame:

$$sf(PI) = PI \mod SF \quad (5)$$

where $X \mod Y$ is the modulo operation. Let's suppose that the vehicle has to transmit a packet that occupies $N_p$ sub-channels (with $N_p \geq 1$). The first sub-channel that the vehicle with *PosIndex PI* selects in sub-frame $sf(PI)$ is the sub-channel numbered $sc(PI)$ with:

$$sc(PI) = 2 \cdot R(PI) \cdot N_p \mod SC + \left\lfloor \frac{2 \cdot R(PI) \cdot N_p}{SC} \right\rfloor \cdot N_p \quad (6)$$

and

$$R(PI) = \left\lfloor \frac{PI}{SF} \right\rfloor \cdot \mod \left\lfloor \frac{SC}{N_p} \right\rfloor \quad (7)$$

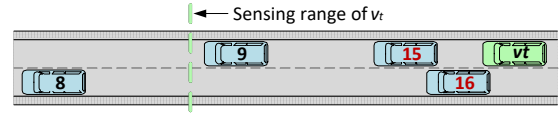

Fig. 3. Illustration of trade-offs in the *PosIndex* estimation.

If $N_p > 1$, the vehicle selects the sub-channels numbered $sc(PI)$ to $sc(PI) + N_p$ -1 in sub-frame $sf(PI)$. If there are more vehicles in the scenario than available sub-channels, vehicles with the same *PosIndex* will share the same sub-channel(s). However, the scheme maximizes the distance between vehicles that share the same sub-channel(s). This minimizes the interference and reduces the risk of packet collisions. It should be noted that the proposed sub-channel selection process is valid for different packet sizes. The modulation and coding scheme (MCS) can be adapted to fit a packet into $N_p$ sub-channels with $N_p$ between 1 and *SC*.

### D. Randomization

The proposed scheme avoids packet collisions between vehicles (even with vehicles outside the sensing range) if the *PosIndex* values are correctly estimated. Most errors in the *PosIndex* estimation are rapidly detected and fixed by the proposed scheme. However, some errors could persist over time and degrade the performance as illustrated in Fig. 4. Let's suppose that vehicles in the left group of Fig. 4.a cannot sense the vehicles on the right group. In this case, vehicles compute their *PosIndex* independently of the *PosIndex* values of the vehicles in the other group. It could then happen that the last vehicle of the left group has the same *PosIndex* as the first vehicle of the right group. When the two groups get closer and vehicles start sensing each other (Fig. 4.b), the transmissions from the two vehicles with the same *PosIndex* will collide. These vehicles will not be able to detect the packet collision that will persist over time. In addition, nearby vehicles will not receive the packets from the two vehicles that collide, so the following vehicle (2[nd] vehicle in the right group) will select their *PosIndex* (Fig. 4.b). This error will propagate (Fig. 4.c) and the transmissions from all vehicles will collide since they all end up selecting the same *PosIndex* value.

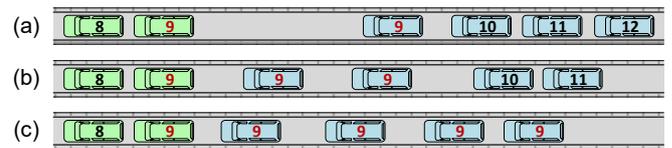

Fig. 4. *PosIndex* error propagation.

To avoid these errors, we introduce some randomization in the proposed scheduling scheme. In particular, each vehicle is requested to transmit one out of $w$ packets in a sub-channel(s) selected randomly. These packets still include the *PosIndex* estimated following the process previously explained. $w$ is randomly chosen between $w_{min}$=5 and $w_{max}$=15 for $\lambda$=10pps, between $w_{min}$=10 and $w_{max}$=30 for $\lambda$=20pps, and between $w_{min}$=25 and $w_{max}$=75 for $\lambda$=50pps. These values have been chosen considering that 3GPP specifications allow C-V2X mode 4 vehicles to use the same sub-channels between 0.5s (equal to $w_{min}/\lambda$) and 1.5s (equal to $w_{max}/\lambda$). We also consider these time limits for our scheme. The probability that a vehicle has to transmit using a randomly selected sub-channel is then:

$$p^{ran} = \left(\frac{w_{min} + w_{max}}{2}\right)^{-1} \quad (8)$$

The random sub-channel(s) is selected from a set of sub-channels that excludes the sub-channel(s) corresponding to the *PosIndex* of the vehicle. This set is chosen to maximize the distance with potentially colliding vehicles, and is referred to as Random Transmission Window. This distance can be maximized if a vehicle with *PosIndex PI* selects the Random Transmission Window around the sub-channel corresponding to the *PosIndex PR*:

$$PR = \begin{cases} PI + \lceil N/2 \rceil - SF/2 & if \quad SC \bmod 2 = 0 \\ PI + \lceil N/2 \rceil & if \quad SC \bmod 2 = 1 \end{cases} \quad (9)$$

Following this equation, the vehicle randomly selects a sub-channel within the set of sub-channels identified by the *PosIndex* values included in the interval [$(PR-M) \bmod N$, $(PR+M) \bmod N$]. In this study, $M$ has been set equal to 5, 2 and 1 for vehicles transmitting 10pps, 20pps and 50pps, respectively. These values have been chosen to guarantee with 99% probability that two vehicles with the same *PosIndex* can detect each other during their random transmissions. This will allow them to detect that they share the same *PosIndex* and that they have to correct them. For example, two vehicles do their random transmission in the same *Pool* with probability $p^{ran}$=0.1 when $\lambda$=10pps. The probability that both vehicles randomly select the same sub-channel is $1/(2M+1)$=0.09 ($M$=5 for $\lambda$=10pps). Vehicles can then detect each other during their random transmissions with 99.1% probability.

The distance between potentially colliding vehicles is maximized if $PR=PI+N/2$. However, this value can generate half-duplex errors with nearby vehicles (e.g. those with *PosIndex* equal to *PI*+1 or *PI*-1) when $SC$ is even. To avoid this, we subtract $SF/2$ in (9) when $SC$ is even. Fig. 5 illustrates an example for a vehicle with *PosIndex* equal to 3, a pool with $N$=80 sub-channels ($SF$=20 and $SC$=4) and $M$ equal to 1 (i.e. 50pps). The sub-channel corresponding to *PI*=3 is shaded in the figure. The sub-channels included in the Random Transmission Window are stripped in Fig. 5, and correspond to *PosIndex* values within the interval [52,54]. The random transmission within this interval can collide with the transmissions from vehicles with *PosIndex* 52 to 54. However, the impact of such collisions is mitigated by the distance between the colliding vehicles. Without the term SF/2 in (9), the Random Transmission Window would correspond to *PosIndex* values from 42 to 44 (sub-frames from 2 to 4), which could cause a half-duplex error between the vehicle with *PI*=3 (when it operates in random mode) and its adjacent vehicles with *PI*=2 and *PI*=4 (when they operate in normal mode).

The proposed scheduling scheme has been presented in all Section III considering scenarios with a single lane and driving direction. However, it is perfectly valid for scenarios with multiple lanes and driving directions since the order in the virtual queue is established based on the vehicles' location. Different driving directions increase the rate at which the *PosIndex* values change. However, Section VI demonstrates that the proposed scheduling scheme can cope with these changes, and significantly outperform sensing-based SPS.

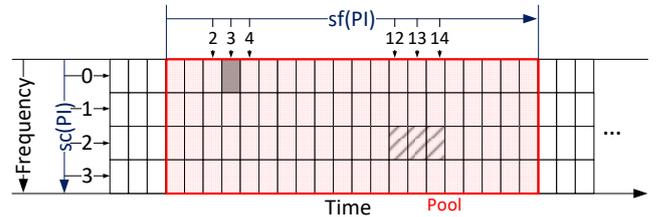

Fig. 5. Random Transmission Window.

## IV. ANALYTICAL PERFORMANCE MODELS

The performance of the proposed scheduling scheme is compared against that achieved with sensing-based SPS. To this aim, this study evaluates analytically and through simulations their PDR (Packet Delivery Ratio) performance as a function of the distance between transmitter and receiver. The PDR is modelled following the approach presented in [19] to evaluate the performance of C-V2X mode 4 using sensing-based SPS. These models are here extended to analytically quantify the C-V2X performance when using the proposed geo-based scheduling scheme. To derive the PDR, we quantify the following four mutually exclusive types of transmission errors present in C-V2X:

1) Errors due to half-duplex transmissions (*HD*). A half-duplex error happens when a receiving vehicle is transmitting a packet in the same sub-frame. The packet is then lost since the receiving vehicle cannot transmit and receive in the same sub-frame. The scheduling scheme has hence an impact on HD errors that depend on the probability that two vehicles select the same sub-frame to transmit their packets.
2) Errors due to a received signal power below the sensing power threshold (*SEN*). A packet cannot be decoded if it is received with a signal power below the sensing power threshold $P_{SEN}$. The probability of suffering an SEN error depends on the distance between transmitter and receiver, the sensing power threshold, the transmission power and the propagation, but not on the scheduling scheme.
3) Errors due to propagation effects (*PRO*). PRO errors are produced when the received signal power of a packet is higher than $P_{SEN}$, but the experienced SNR (Signal to Noise Ratio) is not sufficient to correctly decode the packet. The SNR is computed considering only propagation effects and not interference and packet collisions. PRO errors depend on the same factors as *SEN* errors plus on the MCS used to transmit a packet. In this study, PRO errors exclude *HD* and *SEN* errors.

4) Error due to packet collisions (*COL*). This error is produced when a vehicle transmits on the same sub-channel and sub-frame than another vehicle, and the resulting SINR (Signal to Interference and Noise Ratio) is not sufficient to correctly decode the packet. COL errors depend on the scheduling scheme, the traffic density, the transmission parameters, the distance between transmitter and receiver, and the propagation. In this study, COL errors exclude those quantified in 1), 2) and 3).

We can then assume that a packet is correctly received if none of the four possible errors occur. The PDR is then:

$$PDR(d_{t,r}) = (1-\delta_{SEN}(d_{t,r})) \cdot (1-\delta_{PRO}(d_{t,r})) \cdot (1-\delta_{HD}(d_{t,r})) \cdot (1-\delta_{COL}(d_{t,r})) \quad (10)$$

where $\delta_{HD}$, $\delta_{SEN}$, $\delta_{PRO}$ and $\delta_{COL}$ represent, respectively, the probability of not correctly receiving a packet due to HD, SEN, PRO and COL errors. $d_{t,r}$ represents the distance between the transmitting ($v_t$) and receiving vehicles ($v_r$).

The PDR is here analytically derived for a multi-lane highway scenario with a traffic density of $\beta$ vehicles per meter (i.e. average distance between vehicles of $1/\beta$ meters). We define $\Delta PI_{t,r}$ as the difference between the *PosIndex* of vehicles $v_t$ and $v_r$. This difference is related to the distance between the two vehicles, the traffic density and the number of sub-channels in the *Pool*:

$$\Delta PI_{t,r} = \|\beta \cdot d_{t,r}\| \bmod N \quad (11)$$

where the operation $\|x\|$ represents the nearest integer of $x$.

All vehicles transmit $\lambda$ packets per second with power $P_t$ in a 10MHz channel at 5.9GHz. Following the 3GPP traffic model [20], each vehicle periodically generates beacons or packets. One out of five packets contains 300 bytes of data (referred to as LF or Low Frequency), while the other four contain 190 bytes (HF or High Frequency).

### A. HD errors

To compute $\delta_{HD}$, we need to differentiate when the vehicle is transmitting in the sub-channel corresponding to its *PosIndex* (normal mode) and when it is using the random sub-channel (random mode). The probability that vehicle $v_r$ cannot receive a packet due to the HD effect can be approximated as:

$$\delta_{HD}(d_{t,r}) = p^{ran} \cdot \delta_{HD}^{ran}(d_{t,r}) + p^{nor} \cdot \delta_{HD}^{nor}(d_{t,r}) \quad (12)$$

where $\delta_{HD}^{ran}(d_{t,r})$ and $\delta_{HD}^{nor}(d_{t,r})$ represent the probability that $v_r$ cannot receive a packet due to the half-duplex effect when $v_t$ operates in random and normal mode, respectively. $p^{nor}$ is the probability that $v_t$ operates in normal mode. Let's consider the example in Fig. 5 where $v_t$ has a *PosIndex* equal to 3 and selects the sub-channel in grey when operating under normal mode. In this case, vehicles operating under normal mode and transmitting in the same sub-frame as $v_t$ will not receive the packet from $v_t$. Their probability of HD error is then equal to the probability that they are in normal mode, i.e. $p^{nor} = 1 - p^{ran}$.

An HD error can also occur when $v_t$ is operating under normal mode (*PosIndex* of 3 and the grey sub-channel in Fig. 5) and the receiving vehicles that normally operate under sub-frames 12, 13 and 14 (Fig. 5) operate under random mode. Following (9), these vehicles can select a sub-channel in the same sub-frame as the grey sub-channel when they operate in random mode. In this case, the probability of HD error is equal to the probability that the receiving vehicle is in random mode multiplied by the probability that it selects the same sub-frame as $v_t$ when $v_t$ operates under normal mode (i.e. sub-frame 3 in Fig. 5). In this case, the probability of HD error is equal to $p^{ran}/(2M+1)$. The probability of HD error when $v_t$ operates under normal mode can then be expressed as:

$$\delta_{HD}^{nor}(d_{t,r}) = \begin{cases} p^{nor} & \text{if} \quad \Delta PI_{t,r} \in \{0, K/2, K, 3K/2\} \\ p^{ran}/(2M+1) & \text{if} \quad \Delta PI_{t,r} \in R_{K/4} \cup R_{3K/4} \cup R_{5K/4} \cup R_{7K/4} \\ 0 & \text{else} \end{cases} \quad (13)$$

Where $K = N/\lfloor SC/2 \rfloor$ and $R_X$ is the set of $2M+1$ *PosIndex* values centered at $X$:

$$R_X = \{X-M, \ldots, X+M\} \quad (14)$$

Under random mode, $v_t$ selects a sub-channel and sub-frame following (5), (6) and (9). In Fig. 5, the set of candidate sub-channels include sub-channel 2 in sub-frames 12, 13 and 14. The probability that vehicles in these sub-frames experience an HD error is equal to the probability that they operate in normal mode multiplied by the probability that $v_t$ selects their sub-frame for its random transmission. This probability of HD error is equal to $p^{nor}/(2M+1)$. Similarly, an HD error can occur if vehicles with a normal transmission in sub-frames 2, 3 and 4 operate under random mode, and select the same sub-frame as $v_t$ when $v_t$ is operating under random mode. This probability of HD error is equal to $p^{ran}/(2M+1)$. The probability of HD error when $v_t$ operates under random mode is then:

$$\delta_{HD}^{ran}(d_{t,r}) = \begin{cases} p^{nor}/(2M+1) & \text{if} \quad \Delta PI_{t,r} \in R_{K/4} \cup R_{3K/4} \cup R_{5K/4} \cup R_{7K/4} \\ p^{ran}/(2M+1) & \text{if} \quad \Delta PI_{t,r} \in R_0 \cup R_{K/2} \cup R_K \cup R_{3K/2} \\ 0 & \text{else} \end{cases} \quad (15)$$

### B. SEN errors

*SEN* errors are not influenced by the scheduling scheme. The probability of error because a packet is received with a signal power below the sensing power threshold $P_{SEN}$ is then the same whether using the sensing-based SPS scheme or our proposal. This probability was derived in [19] and is equal to:

$$\delta_{SEN}(d_{t,r}) = \frac{1}{2}\left(1 - erf\left(\frac{P_t - PL(d_{t,r}) - P_{SEN}}{\sigma\sqrt{2}}\right)\right) \quad (16)$$

where $erf$ is the well-known error function, $d_{t,r}$ is the distance between transmitter and receiver, $P_t$ is the transmission power, $PL(d_{t,r})$ is the pathloss at $d_{t,r}$, and $\sigma$ is the variance of the shadowing (*SH*) that is modeled as a log-normal random distribution with zero mean. Readers are referred to [19] for details on how this probability was computed.

## C. PRO errors

The probability that a packet is lost due to propagation effects does not depend on the scheduling scheme. It depends on the SNR of the received packet and the PHY layer performance of the receiver. The PHY layer performance depends on the size of the packet and the MCS used to transmit it. [19] presented an analytical model for $\delta_{PRO}$ considering all packets of equal size and using the same MCS. This paper extends this model to consider LF and HF packets encoded with different MCS. In this case, the probability that a packet is lost due to propagation effects can be expressed as:

$$\delta_{PRO}(d_{t,r}) = p^{HF} \cdot \delta_{PRO}^{HF}(d_{t,r}) + p^{LF} \cdot \delta_{PRO}^{LF}(d_{t,r}) \quad (17)$$

where $p^{LF}(=1/5)$ and $p^{HF}(=4/5)$ are the probability to transmit an LF or HF packet. $\delta_{PRO}^{HF}(d_{t,r})$ and $\delta_{PRO}^{LF}(d_{t,r})$ represent the probabilities that an HF or LF packet is lost due to propagation effects, respectively. The *SNR* at a receiver can be expressed (in dB) as a random variable:

$$SNR(d_{t,r}) = P_r(d_{t,r}) - N_0 = P_t - PL(d_{t,r}) - SH - N_0 \quad (18)$$

where $N_0$ is the noise power. The pathloss is constant at a fixed distance $d_{t,r}$, so the SNR follows the same distribution as the shadowing *SH* but with a mean value equal to $P_t$ - *PL* - $N_0$.

We use the link level Look-Up Tables (LUTs) presented in [22] to model the C-V2X PHY layer performance. These LUTs relate the Block Error Rate (BLER) to the *SNR* for a packet size, MCS, scenario (highway or urban), and relative speed between transmitter and receiver. The probability that an HF packet is lost due to propagation effects is:

$$\delta_{PRO}^{HF}(d_{t,r}) = \sum_{s=-\infty}^{+\infty} BLER^{HF}(s) \cdot f_{SNR|P_r>P_{SEN},d_{t,r}}(s) \quad (19)$$

where

$$f_{SNR|P_r>P_{SEN},d_{t,r}}(s) = \begin{cases} \dfrac{f_{SNR,d_{t,r}}(s)}{1-\delta_{SEN}} & \text{if } P_r > P_{SEN} \\ 0 & \text{if } P_r \leq P_{SEN} \end{cases} \quad (20)$$

$BLER^{HF}(s)$ represents the BLER for an SNR equal to *s* and the MCS utilized to transmit the HF packets. $f_{SNR,d_{t,r}}(s)$ represents the PDF of the *SNR*. $f_{SNR|P_r>P_{SEN},d_{t,r}}(s)$ is the PDF of the SNR at $d_{t,r}$ for SNR values with $P_r>P_{SEN}$. The objective of this term is to omit packets with a received signal power lower than $P_{SEN}$ since they have already been taken into account in (16). The PDF of the *SNR* needs to be normalized by 1- $\delta_{SEN}$ in (20) so that the integral of this equation between -∞ and +∞ is 1, and the probability $\delta_{PRO}^{HF}$ is between 0 and 1.

$\delta_{PRO}^{LF}$ is derived similarly, and can be expressed as:

$$\delta_{PRO}^{LF}(d_{t,r}) = \sum_{s=-\infty}^{+\infty} BLER^{LF}(s) \cdot f_{SNR|P_r>P_{SEN},d_{t,r}}(s) \quad (21)$$

## D. COL errors

Errors due to packet collisions are produced when an interfering vehicle $v_i$ transmits on the same sub-frame and sub-channel than the transmitting vehicle $v_t$, and the interference generated results in an SINR that is not sufficient for the receiving vehicle $v_r$ to correctly decode the packet. This error depends on the scheduling scheme, and hence it is necessary to derive the analytical expression for our proposal. The probability of error due to collisions also depends on the packet size, and can hence be expressed as:

$$\delta_{COL}(d_{t,r}) = p^{HF} \cdot \delta_{COL}^{HF}(d_{t,r}) + p^{LF} \cdot \delta_{COL}^{LF}(d_{t,r}) \quad (22)$$

where $\delta_{COL}^{HF}(d_{t,r})$ and $\delta_{COL}^{LF}(d_{t,r})$ represent the probability of packet loss due to collisions when the transmitting vehicle transmits a HF or a LF packet, respectively. These probabilities can be expressed as a function of the probability that the transmitting vehicle is in normal or random mode, and the probabilities that packet collisions occur when operating under each mode and when transmitting HF or LF packets:

$$\delta_{COL}^{HF}(d_{t,r}) = p^{nor} \cdot \delta_{COL}^{nor,HF}(d_{t,r}) + p^{ran} \cdot \delta_{COL}^{ran,HF}(d_{t,r}) \quad (23)$$

$$\delta_{COL}^{LF}(d_{t,r}) = p^{nor} \cdot \delta_{COL}^{nor,LF}(d_{t,r}) + p^{ran} \cdot \delta_{COL}^{ran,LF}(d_{t,r}) \quad (24)$$

where:

$$\delta_{COL}^{nor,HF}(d_{t,r}) = 1 - \prod_i \left(1 - \delta_{COL}^{nor,HF,i}(d_{t,r},d_{t,i},d_{i,r})\right) \quad (25)$$

$$\delta_{COL}^{nor,LF}(d_{t,r}) = 1 - \prod_i \left(1 - \delta_{COL}^{nor,LF,i}(d_{t,r},d_{t,i},d_{i,r})\right) \quad (26)$$

$$\delta_{COL}^{ran,HF}(d_{t,r}) = 1 - \prod_i \left(1 - \delta_{COL}^{ran,HF,i}(d_{t,r},d_{t,i},d_{i,r})\right) \quad (27)$$

$$\delta_{COL}^{ran,LF}(d_{t,r}) = 1 - \prod_i \left(1 - \delta_{COL}^{ran,LF,i}(d_{t,r},d_{t,i},d_{i,r})\right) \quad (28)$$

Eq. (25) to (28) are expressed as a function of the interfering vehicle $v_i$ and its distance to the transmitting and receiving vehicles. $v_i$ can provoke a COL error if $v_t$ and $v_i$ simultaneously transmit using the same sub-channel, and the interference generated by $v_i$ is such that the packet is lost due to the SINR. The previous equations can be expressed as:

$$\delta_{COL}^{nor,HF,i}(d_{t,r},d_{t,i},d_{i,r}) = p_{SIM}^{nor,HF}(d_{t,i}) \cdot p_{INT}^{HF}(d_{t,r},d_{i,r}) \quad (29)$$

$$\delta_{COL}^{nor,LF,i}(d_{t,r},d_{t,i},d_{i,r}) = p_{SIM}^{nor,LF}(d_{t,i}) \cdot p_{INT}^{LF}(d_{t,r},d_{i,r}) \quad (30)$$

$$\delta_{COL}^{ran,HF,i}(d_{t,r},d_{t,i},d_{i,r}) = p_{SIM}^{ran,HF}(d_{t,i}) \cdot p_{INT}^{HF}(d_{t,r},d_{i,r}) \quad (31)$$

$$\delta_{COL}^{ran,LF,i}(d_{t,r},d_{t,i},d_{i,r}) = p_{SIM}^{ran,LF}(d_{t,i}) \cdot p_{INT}^{LF}(d_{t,r},d_{i,r}) \quad (32)$$

$p_{SIM}^{nor,HF}(d_{t,i})$ in eq. (29) represents the probability that $v_t$ and $v_i$ simultaneously transmit using the same sub-channel when $v_t$ is transmitting an HF packet under normal mode. $p_{INT}^{HF}(d_{t,r},d_{i,r})$ is the probability that the interference generated by $v_i$ on $v_r$ is higher than a threshold that would provoke that if $v_t$ and $v_i$ simultaneously transmit on the same sub-channel, $v_r$ cannot correctly decode the packet. Terms in eq. (30), (31) and (32) are analogous but for LF packets and/or transmissions in random mode. These probabilities are next computed.

*D1. Probability that interference is higher than threshold*

We assume that the effect of the interference from $v_i$ to $v_r$ is equivalent to additional noise. The *SINR* at $v_r$ can then be expressed in dB as:

$$SINR(d_{t,r}, d_{i,r}) = P_r(d_{t,r}) - P_i(d_{i,r}) - N_0 \quad (33)$$

where $P_i$ is the signal power received by $v_r$ from $v_i$. *SINR* is hence a random variable resulting from the sum of two random variables ($P_r$ and $P_i$). The PDF of the *SINR* can hence be obtained from the cross correlation of the PDF of $P_r$ and $P_i$. The probability that $v_r$ incorrectly receives a HF packet due to low *SINR* (i.e. low $P_r$ and/or high $P_i$) is then:

$$p_{SINR}^{HF}(d_{t,r}, d_{i,r}) = \sum_{s=-\infty}^{+\infty} BLER^{HF}(s) \cdot f_{SINR|P_r > P_{SEN}, d_{t,r}, d_{i,r}}(s) \quad (34)$$

Equation (34) includes the HF packets incorrectly received due to propagation effects. These packets were already considered in $\delta_{PRO}^{HF}(d_{t,r})$. To only consider those HF packets lost due to collisions, we perform the following normalization:

$$p_{INT}^{HF}(d_{t,r}, d_{i,r}) = \frac{p_{SINR}^{HF}(d_{t,r}, d_{i,r}) - \delta_{PRO}^{HF}(d_{t,r})}{1 - \delta_{PRO}^{HF}(d_{t,r})} \quad (35)$$

where $\delta_{PRO}^{HF}(d_{t,r})$ is obtained from (19). The same LUTs used in (19) are used in (34) to estimate the BLER in $BLER^{HF}(s)$.

The equations for LF packets are analogous to those of HF packets, and are obtained replacing the HF-dependent variables in (34) and (35) by the LF ones:

$$p_{SINR}^{LF}(d_{t,r}, d_{i,r}) = \sum_{s=-\infty}^{+\infty} BL^{LF}(s) \cdot f_{SINR|P_r > P_{SEN}, d_{t,r}, d_{i,r}}(s) \quad (36)$$

$$p_{INT}^{LF}(d_{t,r}, d_{i,r}) = \frac{p_{SINR}^{LF}(d_{t,r}, d_{i,r}) - \delta_{PRO}^{LF}(d_{t,r})}{1 - \delta_{PRO}^{LF}(d_{t,r})} \quad (37)$$

*D2. Probability that $v_t$ and $v_i$ simultaneously transmit on the same sub-channel*

The probability $p_{SIM}$ that $v_t$ and $v_i$ transmit simultaneously in the same sub-frame and sub-channel depends on whether they operate in normal or random mode, and on whether they transmit a HF or LF packet. When $v_t$ is in normal mode and transmits a HF packet, it can receive interference from vehicles $v_i$ operating in normal mode and using the same *PosIndex* (i.e. $\Delta PI_{t,i}=0$). This situation can occur with probability $p^{nor}$. $v_t$ can also receive interference from a vehicle $v_i$ that transmits an LF packet in normal mode with $\Delta PI_{t,i}=K$. This is case since we consider that LF packets occupy twice the number of sub-channels of HF packets. The probability that this situation occurs is equal to the probability that $v_i$ is in normal mode and transmits an LF packet ($p^{nor} \cdot p^{LF}$). $v_t$ can also receive interference from vehicles $v_i$ operating under random mode. More specifically, when $v_t$ transmits an HF packet under normal mode, it can be interfered with probability $p^{ran}/(2M+1)$ by vehicles $v_i$ operating under random mode and with $\Delta PI_{t,i} \in R_{5K/4}$. In this case, $p^{ran}/(2M+1)$ represents the probability that $v_i$ is in random mode and selects the same sub-frame and sub-channel than $v_t$ (i.e. $1/(2M+1)$). Similarly, $v_t$ can also receive interference from vehicles $v_i$ that transmit an LF packet in random mode with $\Delta PI_{t,i} \in R_{K/4}$. This situation occurs with probability $p^{ran} \cdot p^{LF}/(2M+1)$ since interfering vehicles $v_i$ have to be in random mode and transmit an LF packet. The probability that $v_t$ is interfered when transmitting an HF packet in normal mode can then be expressed as:

$$p_{SIM}^{nor,HF}(d_{t,i}) = \begin{cases} p^{nor} & if \quad \Delta PI_{t,i} = 0 \\ p^{nor} \cdot p^{LF} & if \quad \Delta PI_{t,i} = K \\ p^{ran}/(2M+1) & if \quad \Delta PI_{t,i} \in R_{5K/4} \\ p^{ran} \cdot p^{LF}/(2M+1) & if \quad \Delta PI_{t,i} \in R_{K/4} \\ 0 & else \end{cases} \quad (38)$$

The process to compute the probability that $v_t$ is interfered when transmitting an LF packet in normal mode is analogous to that of HF packets. The main differences are that:

- $v_i$ operating in normal mode and with $\Delta PI_{t,i}=K$ can interfere $v_t$ when it transmits an HF packet. The probability that this happens is $p^{nor}$ rather than $p^{nor} \cdot p^{LF}$.
- $v_i$ operating in random mode and with $\Delta PI_{t,i} \in R_{K/4}$ can interfere $v_t$ with probability $p^{ran}/(2M+1)$ since it can interfere whether transmitting an HF or LF packet.

The probability that $v_t$ is interfered when transmitting an LF packet in normal mode can then be expressed as:

$$p_{SIM}^{nor,LF}(d_{t,i}) = \begin{cases} p^{nor} & if \quad \Delta PI_{t,i} \in \{0, K\} \\ p^{ran}/(2M+1) & if \quad \Delta PI_{t,i} \in R_{K/4} \cup R_{5K/4} \\ 0 & else \end{cases} \quad (39)$$

The potentially interfering vehicles are different when $v_t$ is in random mode. Under this mode, the vehicles $v_i$ that can interfere $v_t$ when $v_t$ transmits an HF packet are:

- Vehicles $v_i$ with a *PosIndex* such that $\Delta PI_{t,i} \in R_{7K/4}$. This interference can occur with probability $p^{nor}/(2M+1)$. This is the case because $v_i$ has to operate under normal mode and it does not depend on the type of packet it transmits (both LF and HF interfere in this case). Additionally, $v_t$ has to randomly select the same sub-channel as $v_i$ (there are $2M+1$ possible sub-channels).
- Vehicles $v_i$ with a *PosIndex* such that $\Delta PI_{t,i} \in R_{3K/4}$. The interference can occur if vehicles $v_i$ transmit a LF packet in normal mode, which happens with probability $p^{nor} \cdot p^{LF}/(2M+1)$.
- Vehicles $v_i$ operating in random mode and with a *PosIndex* such that $\Delta PI_{t,i} \in R_0$. In this case, the probability of interference is $p^{ran}/(2M+1)$.
- Vehicles $v_i$ transmitting an LF packet in random mode and with a *PosIndex* such that $\Delta PI_{t,i} \in R_K$. This situation can occur with probability $p^{ran} \cdot p^{LF}/(2M+1)$.

The probability that an interfering vehicle $v_i$ simultaneously transmits in the same sub-channel than $v_t$ when $v_t$ is in random mode and transmits a HF packet can then be expressed as:

$$p_{SIM}^{ran,HF}(d_{t,i}) = \begin{cases} p^{nor}/(2M+1) & if \quad \Delta PI_{t,i} \in R_{7K/4} \\ p^{nor} \cdot p^{LF}/(2M+1) & if \quad \Delta PI_{t,i} \in R_{3K/4} \\ p^{ran}/(2M+1) & if \quad \Delta PI_{t,i} \in R_0 \\ p^{ran} \cdot p^{LF}/(2M+1) & if \quad \Delta PI_{t,i} \in R_K \\ 0 & else \end{cases} \quad (40)$$

The potentially interfering vehicles $v_i$ are the same when $v_t$

transmits a LF packet under random mode than when it transmits a HF packet. Equation (40) can then be adapted to compute the probability that $v_t$ and $v_i$ simultaneously transmit on the same sub-channel when $v_t$ is in random mode and transmits an LF packet. In this case, vehicles $v_i$ could interfere $v_t$ irrespective of whether they transmit a LF or a HF packet. As a result, (40) needs to be modified by removing $p^{LF}$ in order to compute the probability that $v_i$ interferes $v_t$ when $v_t$ is in random mode and transmits a LF packet:

$$p_{SIM}^{ran,LF}(d_{t,i}) = \begin{cases} p^{nor}/(2M+1) & if \quad \Delta PI_{t,i} \in R_{3K/4} \cup R_{7K/4} \\ p^{ran}/(2M+1) & if \quad \Delta PI_{t,i} \in R_0 \cup R_K \\ 0 & else \end{cases} \quad (41)$$

The probability of packet loss due to collisions ($\delta_{COL}$) is then computed with (22) and (23)-(41). Finally, the PDR is computed using (10) where $\delta_{HD}$, $\delta_{SEN}$, $\delta_{PRO}$ and $\delta_{COL}$ are obtained from (12), (16), (17) and (22), respectively.

## V. EVALUATION SCENARIO

The scheduling schemes have been evaluated under a Highway scenario following 3GPP recommendations in [20]. The scenario implements a 5km highway segment with 6 lanes (3 lanes in each direction). The maximum speed is 140km/h for a traffic density of 60veh/km and 70km/h for 120veh/km. The mobility is modeled with the traffic simulator SUMO.

The scheduling schemes are evaluated considering C-V2X mode 4 communications. Such communications are simulated in VEINS using a C-V2X interface implemented by the authors following the 3GPP specifications in [20]. The C-V2X mode 4 interface (using sensing-based SPS) was validated against an analytical model in [19]. The proposed scheduling scheme has been implemented on the same simulator. The simulator models propagation using the WINNER+ B1 model recommended by 3GPP in [20]. The model implements a log-distance pathloss model, and models the shadowing using a log-normal random distribution with a standard deviation of 3dB. It also implements the shadowing correlation as specified in [20]. The simulator models the C-V2X PHY layer performance using the BLER-SNR curves from [22] that include the fast fading effect. Following the 3GPP guidelines [20], we assume perfect time and frequency synchronization at sub-frame and sub-carrier levels and a noise figure of 9dB. The simulator implements the In-Band Emission model defined by the 3GPP in [20]. All vehicles transmit beacons at 23dBm in a dedicated 10MHz channel at 5.9GHz. The channel is divided into 4 sub-channels of 12 RBs each. Beacons are periodically generated following the traffic model specified by the 3GPP in [20]. The model specifies that one out of five beacons contains 300 bytes of data (LF beacon), while the remaining four contain 190 bytes (HF beacons). LFs are transmitted using MCS 7, and require 2 sub-channels. HFs are transmitted using MCS 9, and require 1 sub-channel.

## VI. RESULTS

### A. Validation

Fig. 6 compares the PDR obtained with the proposed scheduling scheme analytically (using the model in Section III) and through simulations. The simulation platform implements the complete geo-based scheduling scheme as described in Section V. This implementation provides an accurate estimate of the network performance that can be obtained with the proposed scheme under realistic mobility conditions. On the other hand, the analytical model provides an estimate of the performance that can be achieved when vehicles are perfectly ordered. Fig. 6 shows that both PDR curves are very similar which validates the analytical model and the implementation at the simulation platform. The small differences observed in Fig. 6 are mainly due to In-Band Emissions and to a small percentage of ordering errors in the simulations. The simulation platform models In-Band Emissions, and such emissions suppose some extra collision errors due to the near-far effect. The high mobility of vehicles during the simulations also generates a small percentage of ordering errors that contribute to the small differences between the analytical and simulation PDRs in Fig. 6. A correct operation of the proposed scheduling scheme requires that the *PosIndex* of a vehicle is equal to the *PosIndex*+1 of the preceding vehicle in the scenario. 99.4% and 99.2% of the vehicles in the simulations guaranteed this condition for traffic densities of 60 and 120 vehicles per km, respectively. The remaining vehicles select an incorrect *PosIndex* value. However, a packet collision between two vehicles only occurs if two vehicles select the same *PosIndex*. In the simulated scenarios, only 0.46% and 0.61% of the vehicles selected at a given moment the same *PosIndex* value as their preceding vehicle for traffic densities of 60 and 120 vehicles per km respectively. Their transmissions then collide, and these collisions contribute to the small differences observed in Fig. 6 between the analytical and simulated PDRs. This type of errors can occur when two nearby vehicles detect for a short period of time different vehicles close to their sensing range. In this case, they establish different virtual queues until they both detect the same vehicles.

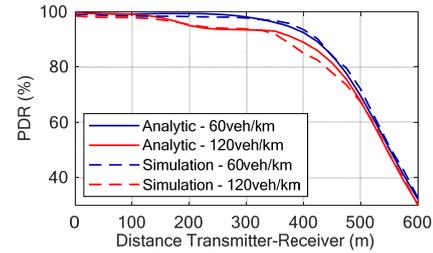

Fig. 6. PDR as a function of the distance between transmitter and receiver. Vehicles transmit 10pps.

Fig. 7 represents an example of the percentage of vehicles that change their position in the queue within a *Pool*, and the percentage of vehicles that incorrectly select their *PosIndex*. The figure clearly shows that only a very small percentage of vehicles incorrectly select their *PosIndex* despite the highly dynamic scenario (on average, around 20% of the vehicles in the scenario change their position in the queue within 100ms). This demonstrates that the geo-based scheduling scheme is robust, and can correctly handle continuous changes of the vehicles' position within the queue.

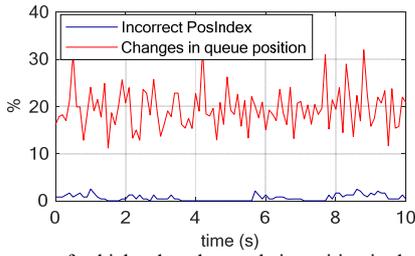

Fig. 7. Percentage of vehicles that change their position in the queue within a *Pool*, and percentage of vehicles that incorrectly select their *PosIndex*. Scenario with 120veh/km and 10pps. Similar trends observed for 60veh/km.

### B. Comparative analysis

Fig. 8 compares the PDR obtained with the proposed geo-based scheduling scheme to that obtained with the sensing-based SPS scheme defined in 3GPP. The comparison is done for simulations with a traffic density of 60 and 120 vehicles/km. Fig. 8 demonstrates that the geo-based scheduling scheme significantly outperforms the sensing-based scheduling algorithm. The figure also shows that our scheduling proposal better copes with an increase in the channel load, and its gains over the sensing-based SPS scheme augment with the traffic density. This is the case because our proposal better organizes the use of sub-channels and hence reduces the risk of packet collisions. This risk increases with the density for the sensing-based SPS scheme as illustrated in Fig. 9. This figure represents the percentage of packets incorrectly received per type of transmission error in C-V2X as a function of the distance between transmitter and receiver. The figure clearly shows that our proposed scheduling scheme significantly reduces the percentage of packets lost due to packet collisions compared to the standard sensing-based SPS scheme. Packet collisions are the dominant factor for packet loses up to a distance of 450m. This distance is where the hidden-terminal effect is stronger for the sensing-based SPS scheme. At this distance, our proposed geo-based scheduling scheme reduces packet collisions by 67.6% and 75.7% for 60 and 120 vehicles/km, respectively. These gains are due to a better and more efficient assignment of sub-channels. Propagation (SEN+PRO errors in Fig. 9) becomes the dominant factor for packet loses for distances higher than 450m. Both scheduling schemes experience the same percentage of packets lost due to propagation since our proposal does not modify the physical layer of C-V2X mode 4. Fig. 9 does not represent the percentage of packets lost due to HD errors since they are negligible for these scenarios. HD errors become more relevant when vehicles transmit more packets per second. Release 14 introduces support for C-V2X communications at 20pps and 50pps. These rates are expected to sustain some of the enhanced V2X use cases that have been specified under 3GPP Release 15 for connected and automated driving. High packet generation rates can challenge the operation of the sensing-based SPS scheme given the large percentage of packets lost due to packet collisions. This is actually observed when comparing the PDR obtained with the sensing-based SPS scheme in Fig. 8 and Fig. 10. Fig. 10 represents the PDR obtained with the two scheduling schemes under evaluation when considering a packet generation rate of 50pps and a traffic density of 60 vehicles/km. The comparison of Fig. 8 and Fig. 10 clearly shows the strong PDR degradation experienced with the standardized sensing-based SPS scheme when the channel load increases as a result of higher packet generation rates. Fig. 10 clearly shows that the proposed scheduling scheme better copes with the increase in the channel load, and improves the PDR compared with the standardized sensing-based SPS scheme. In particular, the proposed scheduling scheme improves the PDR at short distances, and considerably augments the distance at which a PDR of 0.9 can be guaranteed. This PDR threshold is usually considered by the 3GPP to analyze the performance of V2X communications [2][20]. The proposed scheme significantly improves the PDR at short distances at the expense[4] of a lower PDR between 300 and 400m (although still higher than that achieved with sensing-based SPS). This is due to the Half-Duplex effect explained in Section IV.

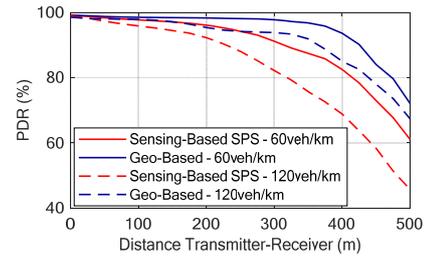

Fig. 8. PDR as a function of the distance between transmitter and receiver for 60 and 120 vehicles/km. Vehicles transmit 10pps.

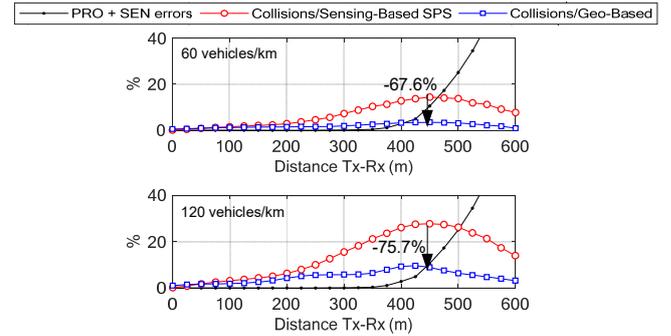

Fig. 9. Percentage of packets lost due to packet collisions and propagation. Vehicles transmit 10pps.

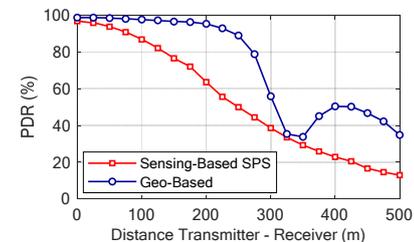

Fig. 10. PDR as a function of the distance between transmitter and receiver for 60 vehicles/km and 50pps.

---

[4] No scheduling scheme can guarantee a high PDR for all distances when the channel load increases since the number of sub-channels is limited. Our proposal has been designed so that the negative effects of such increase are observed for larger rather shorter distances between transmitter and receiver.

Finally, Table II represents the distance at which a PDR equal to 0.9 is guaranteed following the evaluation criteria usually employed in 3GPP. The table depicts the performance obtained with the sensing-based SPS scheme and our geo-based proposal for all the scenarios analyzed in this study. The table also shows the improvement obtained with our proposal compared to the sensing-based SPS scheme. Table II clearly shows that our proposal outperforms the sensing-based SPS scheme in all scenarios. Higher gains are achieved when the channel load increases and the sensing-based SPS scheme suffers from significant packet collisions.

TABLE II. DISTANCE AT WHICH A PDR EQUAL TO 0.9 IS GUARANTEED

| Density (veh/km) | Packet rate (pps) | Sensing-based SPS (m) | Geo-based (m) | Improvement (%) |
|---|---|---|---|---|
| 60 | 10 | 316 | 426 | 34.8 |
|  | 20 | 245 | 407 | 66.1 |
|  | 50 | 80 | 244 | 205 |
| 120 | 10 | 229 | 365 | 59.4 |
|  | 20 | 117 | 283 | 141.9 |
|  | 50 | 31 | 107 | 245.2 |

## VII. CONCLUSIONS

This paper proposes a novel geo-based scheduling scheme for C-V2X communications where vehicles autonomously select their sub-channels. The proposed scheme exploits context information exchanged between vehicles to select the sub-channels. In particular, the scheme takes into account the location of nearby vehicles and their ordering on the road to select the sub-channels. All vehicles follow the same process, and hence implicitly coordinate their sub-channel selection which significantly reduces the number of packet collisions. Analytical models have been presented to quantify the PDR of the proposed scheme, and the probability of packet loss due to the four types of possible transmission errors in C-V2X. Extensive network simulations have also been conducted, and the obtained results demonstrate that the proposed geo-based scheduling scheme reduces packet collisions and significantly increases the C-V2X performance compared to when using the standardized sensing-based SPS scheme. The geo-based scheduling scheme has been evaluated considering C-V2X mode 4. However, it could be applied to C-V2X evolutions where vehicles autonomously select their sub-channels.